\documentclass[a4paper,11pt]{article}
\usepackage{jinstpub} 
\usepackage{lineno}
\usepackage{svg}
\usepackage{ulem}
\modulolinenumbers[5]

\proceeding{Topical Workshop on Electronics for Particle Physics - TWEPP 2024\\
 30 September - 4 October, 2024\\
 Glasgow, Scotland}

\title{\boldmath A Low-Cost, Low-Power Media Converter Solution for Next-Generation Detector Readout Systems}

\author[a,b,1]{A. Perro,\note{Corresponding author.}}
\author[a,c]{M. Vodnik,}
\author[a]{P. Durante}
\affiliation[a]{CERN\\Geneva, Switzerland}
\affiliation[b]{Aix Marseille Univ, CNRS/IN2P3, CPPM\\Marseille, France}
\affiliation[c]{Jozef Stefan Institute\\Ljubljana, Slovenia}
\emailAdd{alberto.perro@cern.ch}

\abstract{High Energy Physics (HEP) data acquisition systems are often built from high-end FPGAs. As such systems scale in the HL-LHC era, severe under-utilization of FPGA transceivers can occur because front-end links prioritize radiation hardness and power consumption over raw data bandwidth. This work evaluates recently introduced low-power, low-cost FPGA devices as an alternative building block for future readout architectures. This study presents the implementation of a readout back-End on FPGA where the front-end protocol is based on the Low-Power GigaBit Transceiver (lpGBT) and the readout protocol is based on 10 Gigabit Ethernet, using the LHCb Run 4 RICH detector as a practical case study.}


\arxivnumber{2410.23173} 

\notoc

\begin{document}
\maketitle
\flushbottom

\section{Introduction}
\label{sec:intro}
\paragraph{LHCb Readout System in Run 3}
The Data Acquisition (DAQ) system of the LHCb experiment \cite{pisani_design_2023} currently employs 11,000 GigaBit Transceiver (GBT) \cite{baron_gbt_2009} links at 4.8 Gbps. The readout system and the Event Builder can handle a cumulative throughput of 32 Tbps. These links are read by custom high-end FPGA boards which aggregate, process, and transfer the data to the Event Builder via PCI Express (PCIe). Each of these FPGA boards has 1.2M Logic Elements, it is capable of handling up to 48 GBT links, and it offers two PCIe 3.0x8 interfaces to the host machine. Of 520 cards employed in LHCb, 445 of them are used for data readout and the rest for control and clock distribution.

The readout cards are hosted in groups of up to three in the Event Building servers. This implies that the data of each event is fragmented over the entire cluster. Data fragments from all sub-detectors have to be collected so that single events can be assembled in the same place to proceed with the reconstruction and selection process. Servers in the Event Building cluster are interconnected by a fast network in order to send and receive data fragments. This network uses the InfiniBand HDR 200 Gbps technology. The choice of this technology over the Ethernet standard was motivated by the performance difference measured for this specific use case at the time of design \cite{krawczyk_feasibility_2020}.

\paragraph{Future DAQs in the HL-LHC era} In the HL-LHC era, the LHCb detector will require major upgrades on the Front-End Electronics (FEE) to take advantage of the higher luminosity available.

Upgraded sub-detectors will require both a faster data rate, provided by Low Power GigaBit Transceiver (lpGBT) technology \cite{paulo_moreira_lpgbt_nodate} reaching up to 10.24 Gbps per link, as well as a higher number of optical links. DAQ systems have to be upgraded accordingly to manage the higher throughput. LHCb estimates a total throughput of 300 Tbps and 30,000 readout links in Run 5 \cite{colombo_online_2024}, 9.4x the throughput and 2.7x the links compared to Run 3.

The DAQ upgrade offers the possibility to design the EB network once again: Ethernet is evolving at a fast pace, with decreasing cost per bandwidth and can be sourced from multiple competing vendors. Unlike InfiniBand, FPGA IP modules are also readily available. Ethernet ASICs, at the time of writing, are capable of switching up to 50 Tbps in a single chip - almost 2x the current LHCb throughput - and throughput is expected to double every two years \cite{margalit_perspective_2021}.

\paragraph{FPGA market trends} The FPGAs currently used in the LHCb readout system provided a sufficient number of transceivers, transceivers' speed, and logic resources at the time of design. However, current high-end FPGAs are moving towards offering fewer transceivers that support very high data bandwidths (up to 112 Gbps) \cite{intel_corporation_agilex_nodate}. FEEs in HEP experiments prioritize radiation hardness and power consumption at the expense of link rates.
This study wants to evaluate the potential of lower-end FPGAs as a cost-effective DAQ platform and to investigate how Ethernet can simplify the DAQ architecture by integrating it on the FPGAs.

\section{Proposed Solution}
\paragraph{Design Principles} 
The proposed design - named \textit{NetGBT} - is guided by the following principles:
\begin{enumerate}
    \item Conversion from custom radiation-hard protocols, such as lpGBT, to a standard network protocol like Ethernet.
    \item Selection of an FPGA featuring an optimal price-to-transceiver ratio, possibly with a high transceiver count.
    \item Support for 10 Gigabit Ethernet (10GbE), with a preference for 25 Gigabit Ethernet (25GbE) where feasible.
    \item Sufficient FPGA resources to decode and aggregate existing and upcoming front-end data protocols.
\end{enumerate}
The use of the Ethernet standard makes the design highly flexible, supporting both direct connections to a network interface card (NIC) for testbenches and small test beam setups, while also scaling efficiently for large-scale deployments. In the latter case, multiple Ethernet uplinks can be aggregated using consumer off-the-shelf (COTS) Ethernet switches.

On top of Ethernet, the Internet Protocol (IP) and the User Datagram Protocol (UDP) Lite are used. These protocols are standard protocols in modern networks and they are supported by every network device.
UDP-Lite is a simple message-oriented protocol based on UDP. It has been designed for use in scenarios where error tolerance is acceptable or built in the encapsulated payload. It finds its use in real-time application such as multimedia streaming. The absence of a full packet checksum simplifies the design of the gateware and streamlines the data processing on the receiver side.

\section{Proof of Concept}
\paragraph{Hardware} The Proof of Concept is based on an off-the-shelf development kit from Opal Kelly \cite{opal_kelly_xem8320_nodate}, which hosts the AMD Artix Ultrascale+ AU25P FPGA. This FPGA features 12 GTY transceivers capable of speeds up to 16.3 Gbps. The baseboard provides access to 4 transceivers via two SYZYGY XCVR connectors \cite{opal_kelly_syzygy_nodate}, 2 transceivers connected to SFP+ cages, and the remaining 2 transceivers routed to SMA connectors. This configuration allows for the handling of up to 4 lpGBT links, which can be converted into 4 10GbE links using two SYZYGY QSFP+ mezzanine cards \cite{opal_kelly_szg-qsfp_nodate}. Additionally, one of the transceivers in the SFP+ cages is allocated for a 1 GbE link, which is used for configuration and control purposes.

However, the selected device does not include 25 Gbps capable transceivers, making it unsuitable for testing the aggregation of multiple lpGBT links onto one 25 GbE output.
\paragraph{Gateware}
The Proof of Concept gateware is designed to handle each lpGBT link separately. The lpGBT links are received using the GTY transceiver hard IP and then decoded using the lpGBT-FPGA core configured for the highest data rate (FEC5, 10.24 Gbps). The decoded words are 224 bit wide and are available at 40 MHz, resulting in a link goodput of 8.96 Gbps.
Once the payload is available, a mixed-width FIFO is used to store the incoming data and do the clock crossing from the 40 MHz lpGBT clock to the 156.25 MHz Ethernet clock.
The data in the FIFO is read in chunks to form multi-word packets: large packets are necessary to make efficient use of the network bandwidth.
Once the packets are formed, the network stack adds the headers for UDP-Lite, IP, and Ethernet. Packets are then forwarded to the 10G/25G Ethernet Subsystem IP which transmits the data to the back-End.

The transmission data path from the FIFO to the MAC header core has been designed using the vendor-independent open source common core library \textit{colibri} \cite{perro_colibri_2024}. Thanks to the use of this library, the gateware has also been ported to a Microchip-based development kit with little adjustments.

Configuration of the system is done via configuration registers which are accessible both via Virtual I/O over JTAG and via a Microblaze Soft Microcontroller, which hosts an MQTT (Message Queuing Telemetry Transport) client to access the registers via the 1 GbE management link. MQTT is a lightweight low-latency protocol designed for communication between devices and is widely used for remote monitoring, telemetry, and data collection applications.

\section{Front-End Emulator}

\paragraph{The Emulator platform}
A font-end emulator has been developed separately to support the development of next generation back-end electronics, by emulating the output of various detector front-end electronics on FPGA gateware.
The emulator is based on the AMD Zynq Ultrascale+ System on Chip, hosted on the ZCU102 evaluation board. The board includes 4 SFP+ cages connected to a Quad of GTH transceivers located on the Zynq chip. Each of the 4 SFP+ ports can be used to emulate one full-speed lpGBT interface. The ZCU102 board also includes an FMC connector which is used to mount a VLDB+ evaluation board, hosting a single lpGBT ASIC chip, which serves as an additional front-end interface.

\paragraph{FastRICH Emulator}
One emulator variant is designed to emulate the data output of 7 instances of the next generation RICH detector readout ASIC: the FastRICH. Two main parameters are available to implement different front-end configurations: number of serial lanes and lane speed. The number of lanes can be chosen from 1 to 4, while lane speed can be set to either 320, 640 or 1280 Mbps. The simulated output data is loaded onto the on-board DDR4 RAM component, and can be continuously streamed via 7 independent channels, each with its Aurora 64B/66B encoder. Depending on the lane configuration, up to 28 serial lanes can be produced, which are distributed over 4 lpGBT optical links. One of the links utilizes the physical lpGBT chip, while the other 3 emulate the lpGBT on FPGA and stream via 3 of the 4 SFP+ ports.

\paragraph{CALO Emulator}
The CALO emulator variant is designed to provide maximum data throughput with just basic data formatting. It is inspired by the LHCb Run 3 Calorimeter readout. It emulates 2 FEE chip instances, each streaming data via 4 channels. The data is formatted into frames with a constant length of 14-bytes, or half of the lpGBT payload. The two instances share 4 lpGBT links, streaming each of the 4 channels on a separate optical fiber. Like in FastRICH emulator, the data is streamed from the RAM component and passed onto the same 4 optical interfaces.

\section{Evaluation}
\label{sec:results}
\begin{figure}[htbp]
    \centering
    \begin{minipage}{0.45\textwidth}
        \centering
\includegraphics[width=\textwidth]{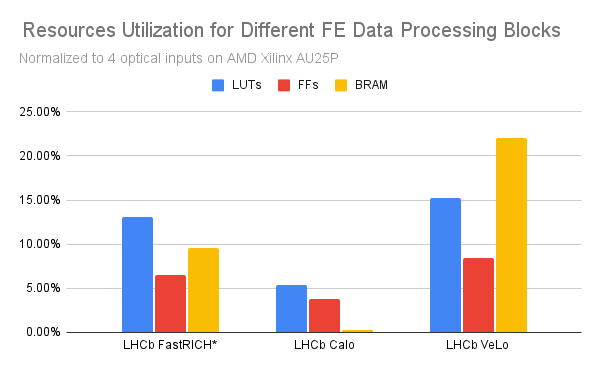}
\caption{Resource Utilization for different data processing blocks, normalized for 4 lpGBT links on the AMD AU25P FPGA. Note that the FastRICH does not implement the full data processing.\label{fig:resources}}
    \end{minipage}\hfill
    \begin{minipage}{0.45\textwidth}
        \centering
        \includegraphics[width=\textwidth]{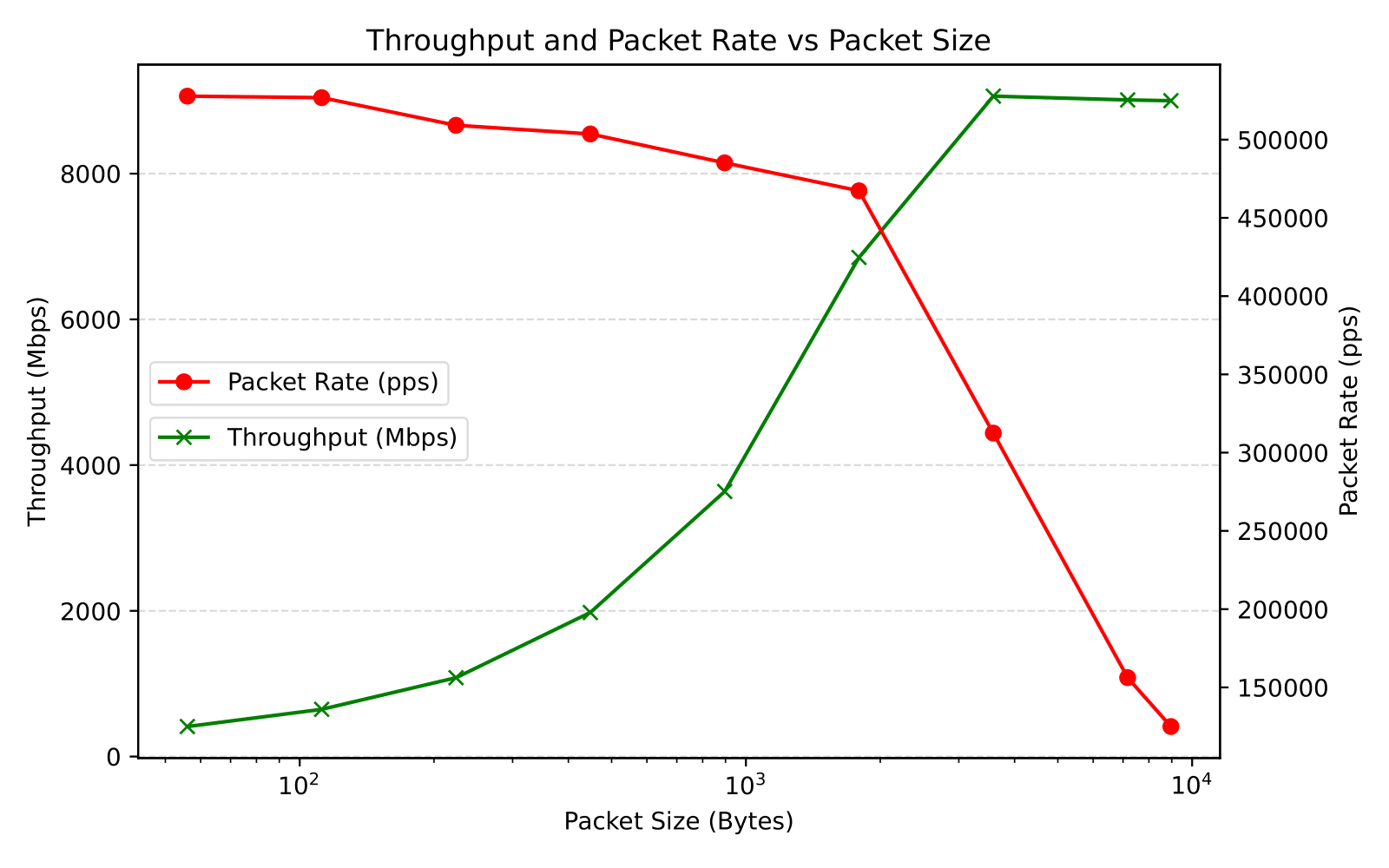} 
\caption{Throughput and packet rate measurements for a single lpGBT link converted to a single 10GbE link.\label{fig:meas}}
    \end{minipage}
\end{figure}

\paragraph{Data Processing} The described gateware uses a small amount of the resources available (Table~\ref{tab:res}).
\begin{table}[htbp]
\centering
\caption{Resource utilization for 4 lpGBT links on AMD AU25P\label{tab:res}}
\smallskip
\begin{tabular}{r|c|c|c|c}
\hline
                 & LUTs  & LUT RAM & FF    & Block RAM \\ \hline
Utilization      & 20056 & 298     & 34500 & 34        \\ \hline
Utilization (\%) & 14.22 & 0.30    & 12.23 & 11.33     \\ \hline
\end{tabular}
\end{table}
This allows the implementation of some FEE-specific data processing on the FPGA, offloading some tasks from the high-level trigger software. Resource utilization results from current and future sub-detector data processing blocks have been measured and normalized to the selected device (Figure~\ref{fig:resources}).

The results indicate that simple front-end data processing involving bit manipulation, error management, and packing (like LHCb CALO) could easily fit on the device. More complex data processing tasks, such as the clustering algorithms employed in the LHCb VeLo, could also fit within the available resources. Additionally, the upcoming FastRICH sub-detector link decoder can also be offloaded onto the device; however, this implementation covers only a portion of the complete data processing as the final dataflow is not yet implemented.

\paragraph{Throughput} Measurements were taken to establish the minimum packet size required to reach the required throughput while avoiding back pressure, packet drops, and subsequent data loss.
The host system is based on a AMD Threadripper 2990WX 32-Core CPU, 64 GB DDR4 RAM, and a Mellanox ConnectX-6 NIC. The OS used is AlmaLinux 9.2 with the 5.14 kernel. The FPGA was connected directly to the NIC via a Direct Attach Copper (DAC) cable.
Some parameters were tuned to ensure best UDP performance in terms of packets per second (PPS):
\begin{itemize}
    \item Enabling Jumbo Packets on the NIC (MTU set to 9000)
    \item Increase the network buffer sizes
    \item Configure \texttt{iptables} to not track and filter packets for the destination UDP port
\end{itemize}
The benchmarks measured the throughput and the packet rate by taking 10 samples of 1 second each to reduce the impact from external factors.

The results show that the point-to-point connection reaches the maximum achievable throughput of ($9064\pm2$) Mbps with the packet size of 3584 B, the correspondent packet rate is ($312,490\pm54$) PPS (Figure \ref{fig:meas}). Therefore, 4 kB is used as the lower bound for packet sizes in this application.

\section{Achievements and Future Directions}
The Proof of Concept demonstrated several promising results, specifically the successful conversion of lpGBT to UDP/IP without back pressure. 
The implementation utilizes few resources, which leaves ample space for additional data processing offload. Furthermore, the solution proved to be cost-effective by using an off-the-shelf mid-range FPGA development kit at a fraction of the cost-per-link compared to high-end FPGAs. The adoption of a standard Ethernet uplink contributes to its flexibility and modularity, specifically the link aggregation using consumer off-the-shelf (COTS) switches.

A second prototype is under development.  This next iteration will utilize a System on Module (SoM) based on the AMD Zynq UltraScale+, featuring 25Gbps-capable transceivers. The new device is designed to support the processing of up to 48 lpGBT links and output through 5 100GbE uplinks. Multiple of these SoMs could be housed in a 1U rack-mounted enclosure, optimizing space efficiency.
\paragraph{Cost Evaluation}

The next iteration of the proposed solution will support 48 lpGBT links. The selected System-on-Module (SoM) is the Enclustra XRU90, priced at \$4,123~\cite{enclustra_gmbh_andromeda_2024} for a single unit, with potential cost reductions at higher volumes. For optical transceivers, a 12-link Samtec FireFly\texttrademark~module is listed at \$920 from the distributor~\cite{mouser_ecuo-y12-14-040-0-1-1-2-21_2024}, resulting in a cost of \$77 per link. By comparison, a COTS Ethernet transceiver is priced at \$40~\cite{sfpcablescom_40gbase-sr4_2024}, resulting in \$10 per link, suggesting the realistic cost for optical links can be around \$43 per link. Using receive-only transceiver modules could further reduce costs to about one-third of this value. Additionally, 100GbE transceivers for the uplink are priced at \$43~\cite{sfpcablescom_100gbase-sr4_2024}.  

The SoM integrates most of the board's cost, and using multiple SoMs on a shared baseboard allows for the distribution of expensive components, such as low-jitter PLLs and voltage regulators. Based on these factors and the pricing of existing SoM baseboards, an additional \$500 per SoM is estimated for baseboard costs.  

This evaluation highlights the cost-effectiveness of the proposed solution, achieving a total cost of \$142 per lpGBT link (Table~\ref{tab:cost}).  

\begin{table}[htbp]
\centering
\begin{tabular}{|r|l|}
\hline

FPGA SoM Cost                         & 4100            \\ \hline
lpGBT Transceivers Cost               & 2000            \\ \hline
100GbE Transceivers Cost              & 200             \\ \hline
PCB Cost                              & 500             \\ \hline
Total Board Cost                      & 6800            \\ \hline
\textbf{Cost per lpGBT link}          & \textbf{142}    \\ \hline
\end{tabular}
\caption{Estimated cost of the NetGBT solution capable of handling 48 lpGBT links. All prices are in \$USD.}
\label{tab:cost}
\end{table}

\acknowledgments

The authors would like to acknowledge the support from the CERN LHCb Online Team, CERN EP R\&D WP 9.3, and ECFA DRD7.5b collaborations.


\bibliographystyle{JHEP.bst}
\bibliography{biblio.bib}


\end{document}